\documentclass[preprint,amsmath,amssymb]{revtex4}

\usepackage{latexsym} 
\usepackage{amssymb}  
\usepackage{amsbsy}   
\usepackage{graphicx}       
\usepackage[dvips]{color}

\newcommand{\al}{\alpha}
\newcommand{\be}{\beta}

\newcommand{\Ga}{\Gamma}
\newcommand{\de}{\delta}

\newcommand{\eps}{\epsilon}

\newcommand{\ka}{\kappa}

\newcommand{\om}{\omega}
\newcommand{\Om}{\Omega}
\newcommand{\Th}{\Theta}

\newcommand{\p}{\partial}
 
\newcommand{\txt}{\textstyle}

\newcommand\Eqn[1]{Eq.~(\ref{#1})}  

\newcommand{\beq}{\begin{equation}}
\newcommand{\eeq}{\end{equation}}
\newcommand{\ba}{\begin{array}}
\newcommand{\bea}{\begin{eqnarray}}
\newcommand{\ea}{\end{array}}
\newcommand{\eea}{\end{eqnarray}}
\newcommand{\bi}{\begin{itemize}}  
\newcommand{\ei}{\end{itemize}}
\newcommand{\ben}{\begin{enumerate}} 
\newcommand{\een}{\end{enumerate}}
\newcommand{\bc}{\begin{center}}
\newcommand{\ec}{\end{center}}
\newcommand{\bl}{\begin{flushleft}}
\newcommand{\el}{\end{flushleft}}
\newcommand{\br}{\begin{flushright}}
\newcommand{\er}{\end{flushright}}

\newcommand\comment[1]{ \hbox{[{\it Comment suppressed here.}\/]} }
\newcommand\hide[1]{}

\newcommand{\mi}{\mathrm{i}}
\newcommand{\dif}{\mathrm{d}}

\newcommand{\skipover}[1]{}
\newcommand{\nn}{\nonumber \\}

\newcommand{\half} {{\txt \frac{1}{2}}}
\newcommand{\third}{{\txt \frac{1}{3}}}

\newcommand{\twothirds}{{\txt \frac{2}{3}}}

\def\lsim{\mathrel{\rlap{\lower4pt\hbox{\hskip1pt$\sim$}}
    \raise1pt\hbox{$<$}}}               
\def\gsim{\mathrel{\rlap{\lower4pt\hbox{\hskip1pt$\sim$}}
    \raise1pt\hbox{$>$}}}                
\usepackage{graphicx}

\numberwithin{equation}{section}

\begin{document}
\title{Viscosity spectral functions of the dilute Fermi gas  
in kinetic theory}

\author{
Matt Braby$^{1,2}$, Jingyi Chao$^1$ and Thomas Sch\"afer$^1$\\
$^1$ Physics Department, North Carolina State University, Raleigh, NC 27695, USA\\
$^2$ Department of Physics and Astronomy, University of North Carolina, Chapel Hill, NC, 27599}

\date{\today}

\begin{titlepage}
\renewcommand{\thepage}{}          

\begin{abstract}
We compute the viscosity spectral function of the dilute Fermi 
gas for different values of the $s$-wave scattering length $a$,
including the unitarity limit $a\to\infty$. We perform the 
calculation  in kinetic theory by studying the response to 
a non-trivial background metric. We find the expected structure 
consisting of a diffusive peak in the transverse shear channel 
and a sound peak in the longitudinal channel. At zero momentum 
the width of the diffusive peak is $\omega_0\simeq (2\varepsilon)
/(3\eta)$ where $\varepsilon$ is the energy density and $\eta$ 
is the shear viscosity. At finite momentum the spectral function
approaches the collisionless limit and the width is of order 
$\omega_0\sim k(T/m)^{1/2}$.

\end{abstract}
\maketitle
\end{titlepage}

\section{Introduction}
\label{sec_intro}

 The dilute Fermi gas at unitarity is a strongly correlated scale
invariant quantum fluid \cite{Bloch:2007,Giorgini:2008}. It provides
a beautiful model system for testing ideas from many different areas 
of physics, including condensed matter physics, nuclear and particle 
physics, and quantum gravity. An important property of the Fermi 
gas at unitarity is nearly perfect fluidity \cite{Schafer:2009dj}.
Experiments with trapped Fermi gases in the vicinity of a Feshbach
resonance indicate that 
\cite{oHara:2002,Schafer:2007pr,Turlapov:2007,Schaefer:2009px,Cao:2010wa,Schaefer:2010dv}
\beq 
\label{eta_s_exp}
\frac{\eta}{s} \lsim 0.5\, \hbar/k_B\, , 
\eeq
where $\eta$ is the shear viscosity and $s$ is the entropy density.
This value is smaller than the viscosity to entropy density of any
known fluid with the possible exception of the quark gluon plasma
\cite{Dusling:2007gi,Romatschke:2007mq,Song:2007ux}. Equ.~(\ref{eta_s_exp})
approaches a bound $\eta/s=\hbar/(4\pi k_B)$ that has been proposed
based on the AdS/CFT correspondence \cite{Kovtun:2004de,Son:2007vk}.
 
 The Kubo formula relates the shear viscosity to the zero momentum,
zero frequency limit of the retarded stress tensor correlation function.
In this work we will study the frequency and momentum dependence of 
this correlation function. We will focus on the regime of high 
temperature and small frequency and momentum where the methods of
kinetic theory are applicable. Our study is motivated by several 
considerations. The first observation is that the frequency and 
momentum dependence of the retarded correlator encodes the time
and distance scales over which the viscous contribution to the 
stress tensor relaxes to the Navier-Stokes expression. These scales
are important for understanding how hydrodynamics breaks down in the 
dilute corona of a trapped Fermi gas \cite{Bruun:2007,Schaefer:2009px}.
Kinetic theory predicts that in the dilute limit the shear viscosity
is independent of density. This leads to a difficulty in the 
analysis of experiments involving scaling flows in which the 
velocity field is linear in the coordinates. The Navier-Stokes 
equation predicts that in this case the viscous stress tensor 
is a function of time, but constant in space. As a consequence 
the dissipative force, which is proportional to gradients of 
the stresses, vanishes and the dissipated heat, which is 
governed by the volume integral of the square of the stress
tensor, is infinite. This problem can be resolved if a finite 
relaxation time for the dissipative stresses is taken into
account. In this case the dissipative stress tensor vanishes 
in the dilute corona, the dissipated heat is finite, and 
there is a viscous force that counteracts the expansion of the 
system. 

 A second motivation is to understand the physics behind nearly 
perfect fluidity in the unitary Fermi, in particular 
the question whether the unitary gas can be described in terms
of quasi-particles. In kinetic theory quasi-particles manifest
themselves in terms of a peak in the spectral function, but 
the quasi-particle peak may disappear in strong coupling. 
The spectral function is accessible through the behavior 
of the euclidean correlation function which can be computed
in imaginary time Monte Carlo simulations. A calculation of 
the spectral function in kinetic theory serves as a useful 
default model for numerical calculations of the euclidean 
correlator. 

 This work extends and complements several recent studies of the 
transport properties of the dilute Fermi gas at unitarity. Calculations
of transport coefficients in the context of kinetic theory can be found 
in \cite{Bruun:2005,Bruun:2006,Rupak:2007vp,Escobedo:2009bh,Braby:2010ec}, 
and a quasi-particle model is discussed in \cite{Guo:2010dc}. A diagrammatic
calculation of the shear viscosity using the Kubo formula is described
in \cite{Enss:2010qh}. This work also studies the spectral function. 
A sum rule for the spectral function of the stress tensor was proved
in \cite{Taylor:2010ju}. Related work on dilute Bose gases can be 
found in \cite{Nikuni:2004}, and first attempts to extend the AdS/CFT 
correspondence to non-relativistic systems are described in
\cite{Son:2008ye,Balasubramanian:2008dm}. Finally, the spectral 
function of the stress tensor in QCD is studied in
\cite{Aarts:2002cc,Teaney:2006nc,Moore:2008ws,Meyer:2008gt,Hong:2010}.

\section{Preliminaries}
\label{sec_pre}

 The retarded correlation function of the stress tensor is defined by
\beq
\label{G_R}
G_R^{ijkl}(\om,{\bf k}) = 
     -i \int \dif t \int \dif {\bf x}\, e^{i\om t - i{\bf k \cdot x}} 
   \Th(t) \langle [\Pi^{ij}(t,{\bf x}), \Pi^{kl}(0,{\bf 0})]\rangle \, . 
\eeq
A careful definition of the stress tensor $\Pi^{\alpha\beta}(t,{\bf x})$
in terms of the Hamiltonian of the dilute Fermi gas was given in 
\cite{Nishida:2007,Enss:2010qh}. In the following, we only need 
the definition of the stress tensor in terms of hydrodynamic and 
kinetic variables, see Sec.~\ref{sec_hydro} and \ref{sec_kin}.
The spectral function is defined by 
\beq
\label{rho_def}
\rho^{ijkl} (\om,{\bf k})
    = -2{\rm Im}\, G^{ijkl}(\om,{\bf k})\, . 
\eeq
In the following, we will focus on the longitudinal and transverse
shear channels $\rho^{xyxy}$, $\rho^{xzxz}$ and $\rho^{zzzz}$ where 
the $z$-direction is the direction defined by the external momentum 
${\bf k}$. We will compute the retarded correlation function using
linear response theory. The external field that couples to the 
stress tensor is the metric $g_{ij}(t,{\bf x})$. A formalism
for coupling an interacting theory of non-relativistic particles 
to an external metric in a way that exhibits a non-relativistic 
version of general coordinate invariance was recently developed
in \cite{Son:2005rv}.  The response to a small perturbation 
$g_{ij}(t,{\bf x})=\delta_{ij}+h_{ij}(t,{\bf x})$ around the 
flat metric is given by 
\beq
\label{G_R_lin}
\de \Pi^{ij} = \frac{\delta \Pi_{ij}^{eq}}{\delta h_{ij}} h^{ij} 
    - \half G_R^{ijkl}h_{kl}
\eeq
where $\Pi^{eq}_{ij}$ is the expectation value of the stress 
tensor in equilibrium. 

\section{Hydrodynamics}
\label{sec_hydro}

 At very low frequency and momentum the retarded correlation function
is governed by hydrodynamics. The hydrodynamic limit provides an 
important constraint for the kinetic description of the system. In
this section we will derive these constraints in the case of 
correlations functions of the stress tensor. We note that hydrodynamics
breaks down for frequencies $\omega\gsim (\eta/n)T$ and momenta $k
\gsim(\eta/n)(mT)^{1/2}$, where $n$ is the density and $T$ is the 
temperature. In the unitary limit and in the vicinity of $T_c$
we have $\eta/n\sim 1$ and the range of validity of hydrodynamics 
is very large. In the weak coupling limit $\eta\sim (mT)^{1/2}/a^2$,
where $a$ is the scattering length, and the range of validity of 
hydrodynamics is much smaller. 

 In order to study linear response theory we consider hydrodynamics
in a curved background $g_{ij}({\bf x},t)$. Non-relativistic fluid
dynamics in a non-trivial metric was studied in \cite{Son:2005tj}.
The continuity equation and the equations of momentum and energy
conservation are given by
\bea
\frac{1}{\sqrt{g}}\partial_{t}\left(\sqrt{g}\rho\right)
   +\nabla_{i}\left(\rho v^{i}\right)  &=& 0\, ,  \nn
\frac{1}{\sqrt{g}}\partial_{t}\left(\sqrt{g}\rho v^{i}\right)
   +\nabla_{j}\Pi^{ji}  &=& 0\, ,  \nn
\frac{1}{\sqrt{g}}\partial_{t}\left(\sqrt{g}\rho s\right)
   +\nabla_{i}\left(\rho sv^{i} -\frac{\kappa}{T}\partial^{i}T\right) 
   &=& \frac{2R}{T} \, , 
\end{eqnarray}
where $\rho=mn$ is the mass density, $v^i$ is the fluid velocity, 
and $g=\det(g_{ij})$. The covariant derivative of a vector field is
given by $\nabla_i v_j = \partial_{i}v_{j}-\Gamma_{ij}^{k}v_{k}$,
where $\Gamma_{ij}^{k}$ is the Christoffel symbol associated with 
$g_{ij}$. The stress tensor is 
\beq
\label{pi_ij}
 \Pi^{ij}=\rho v^{i}v^{j}+Pg^{ij}-\sigma^{ij}, 
\eeq
where the viscous term, $\sigma_{ij}$, is given by
\beq
\label{sig_ij}
\sigma_{ij} = \eta\Big(\nabla_{i}v_{j}+\nabla_{j}v_{i} 
         + \dot{g}_{ij}\Big)
    +\left(\zeta-\frac{2}{3}\eta\right)g_{ij}\left(\nabla_{k}v^{k} 
         + \frac{\dot{g}}{2g}\right)\, . 
\eeq
The dissipative function $R$ is given by
\bea
2R &=& \frac{\eta}{2}\left(\nabla_i v_j + \nabla_j v_i 
   - \twothirds g_{ij} \nabla_k v^k + \dot{g}_{ij} 
   - \third g_{ij} \frac{\dot{g}}{g}\right)^2 \nn
&& + \zeta\left(\nabla_i v^i + \frac{\dot{g}}{2g}\right)^2 
  + \frac{\kappa}{T} \p_i T \p^i T, 
\eea
where $\kappa$ is the thermal conductivity. We are interested in small 
deviations from equilibrium. For this purpose we write $g_{ij} = \de_{ij} 
+ h_{ij}$, and $\rho=\rho_0+\delta\rho$, $P=P_0+\delta P$ etc.~and linearize 
the equations in $h_{ij}$, $v_{i}$, $\delta P$, $\delta s$, $\delta\rho$ and 
$\delta T$. After going to Fourier space we find
\bea
\frac{\mi}{2}\omega\rho_{0}h +\mi\omega\delta\rho
   + \rho_{0}(-\mi k)v_{z}&=& 0\, ,  \nn
\mi\omega\rho_{0}v^{i}+(-\mi k)\delta P \de^{iz}+\mi k\sigma^{iz} &=& 0\, , \nn
\mi\omega\rho_{0}\delta s+\frac{\kappa k^{2}}{T}\delta T &=& 0\, , 
\end{eqnarray}
where $h = {\rm Tr} (h_{ij})$ and we have chosen the direction of 
$k$ to be the $z$-axis. The fluctuations in thermodynamic variables
are related by thermodynamic identities. We will consider $\delta\rho$ 
and $\delta s$ to be the independent variable and use
\bea
\de P &=& \frac{\p P}{\p \rho}\big|_s\de\rho
         +\frac{\p P}{\p s}\big|_{\rho}\de s 
       = c_{s}^{2}\de\rho + \frac{\p P}{\p s}\big|_{\rho}\de s\, ,  \nn
\de T &=&\frac{\p T}{\p \rho}\big|_s\de\rho
         +\frac{\p T}{\p s}\big|_{\rho}\de T 
       = \frac{\p T}{\p \rho}\big|_s\de\rho+\frac{T}{c_V}\de s\, , 
\eea
where $c_{s}$ is the speed of sound and $c_{V}$ is specific heat at 
constant volume.  

 We can now solve for the hydrodynamic variables in terms of the 
external field $h_{ij}$ and compute $\Pi_{ij}$ from \Eqn{pi_ij}.
We then determine $G_R$ using \Eqn{G_R_lin}.  We are particularly
interested in the transverse shear correlators $G_R^{xyxy}$ and 
$G_R^{xzxz}$, as well as the longitudinal correlator $G_R^{zzzz}$.  
We find
\bea
\label{G_xy_hydro}
G_R^{xyxy} &=& -\mi\eta\om \, , \\
\label{G_xz_hydro}
G_R^{xzxz} &=& -\frac{\mi\eta\om^2}{\om-\mi k^2(\eta/\rho_0)}\, ,  \\
\label{G_zz_hydro}
G_R^{zzzz} &=& -\frac{\om^2\rho_0 (c_s^2+\mi\om\al)}
                     {\om^2-c_s^2k^2-\mi\om k^2\al}\, .
\eea
We note that $G_R^{xzxz}$ has a diffusive pole, where $\eta/\rho_0$ 
is the momentum diffusion coefficient, and $G_R^{zzzz}$ has a sound
pole where $\alpha$ is the sound attenuation coefficient, 
\beq
\al\rho_0 = \frac{4}{3}\eta + \zeta + \frac{\ka c_s^2 k^2}{\om^2}
\left(\frac{1}{c_V} - \frac{1}{c_P}\right)\, . 
\eeq 
Finally, we can take the imaginary parts and compute the spectral
functions. In the Kubo limit $k=0$, $\omega\to 0$ we find
\bea
\label{Kubo_T}
\lim_{\om,k\to 0} \frac{\rho_T}{2\om} &=& \eta \, , \\
\label{Kubo_L}
\lim_{\om,k\to 0} \frac{\rho_L}{2\om} &=& \frac{4}{3}\eta + \zeta\, , 
\eea
where $T,L$ labels the transverse and longitudinal components of 
the spectral functions, $\rho_T = \rho^{xyxy},\rho^{xzxz}$ and 
$\rho_L = \rho^{zzzz}$.   

\section{Kinetic Theory}
\label{sec_kin}

 In this section we calculate the spectral function using kinetic
theory. The basic object in kinetic theory is the distribution
function $f({\bf x},{\bf p},t)$ where $f=f_\uparrow=f_\downarrow$. 
The distribution function satisfies the Boltzmann equation. The 
kinetic equation in a curved background can be found by starting 
from the Boltzmann equation in general relativity 
\cite{Andreasson:2002vz,Bruhat:2009}
\beq
\label{boltzmann}
\frac{1}{p^{0}}\left(p^{\mu}\frac{\p}{\p x^{\mu}}
  - \Gamma^{i}_{\alpha\beta}p^{\alpha}p^{\beta}\frac{\p}{\p p^{i}}\right)
    f(t,\mathbf{x},\mathbf{p}) = C[f]\, ,
\eeq
where $i,j,k$ are three-dimensional indices and $\mu,\alpha,\beta$ are 
four-dimensional indices. In the non-relativistic limit $p^0\simeq m$, 
$\Gamma^{i}_{00}\simeq 0$, and $\Gamma^i_{0j}\simeq \frac{1}{2}g^{ik}
\dot{g}_{kj}$ \cite{Son:2005rv}. We get 
\beq
\label{BE_nr}
 \left( \frac{\partial}{\partial t}
                  + \frac{p^i}{m}\frac{\partial}{\partial x^i} 
  - \left(  g^{il}\dot{g}_{lj}p^j
     + \Gamma^{i}_{jk}\frac{p^{j}p^{k}}{m}\right) \frac{\p}{\p p^{i}}\right)
    f(t,\mathbf{x},\mathbf{p}) = C[f]\, ,
\eeq
We consider small deviations from equilibrium and write $f=f_{0}+\delta f$ 
with $f_{0}(\mathbf{p})=f_{0}(p^{i}p^{j}g_{ij}/(2mT))$. We also write 
$g_{ij}=\delta_{ij}+h_{ij}$ and linearize in $h_{ij}$ and $\delta f$. 
We get
\beq
\label{boltzmann_linear}
\left(\p_{t}+\frac{p^i}{m}\p_i\right)\delta f
   + \frac{f_{0}(1-f_{0})}{2mT} p^{i}p^{j}\dot{g}_{ij}
   = C[\delta f]\, . 
\eeq
We will solve the linearized Boltzmann equation by making an 
ansatz for $\delta f$. We go to Fourier space and restrict ourselves 
to quadratic terms in $p$ and write
\beq
\label{del_f_ans}
\de f(\om,{\bf k},{\bf p})=\om f_{0}(1-f_{0}) \frac{p^i p^j}{2mT}
     \frac{\xi_T h^T_{ij}+\xi_L h^L_{ij} }
          {\omega-{\bf v}_{p}\cdot {\bf k}+\mi\epsilon}\, , 
\eeq
where $\xi_{T,L}=\xi_{T,L}(\omega,{\bf k})$ and 
\beq
h^{T}_{ij}=h_{ij}-\frac{1}{3}\delta_{ij}h\, ,
\hspace{0.5cm}
h^{L}_{ij}=\frac{1}{3}\delta_{ij}h\, . 
\eeq
Using this ansatz the LHS of the linearized Boltzmann equation
becomes
\beq
LHS = \frac{\mi\om f_{0}(1-f_{0})}{2mT}
  \left\{ (\xi_T+1) p^i p^j h^T_{ij}
         +(\xi_L+1) p^i p^j h^L_{ij} \right\}\, . 
\eeq
The RHS of the Boltzmann equation involves the linearized collision
integral 
\beq
RHS =   \int\dif\Gamma_{234} \; r(1,2;3,4)
  \frac{\om}{2mT}\left[ g_{ij}({\bf p}_1) + g^{ij}({\bf p}_2) 
  - g^{ij}({\bf p}_3) - g^{ij}({\bf p}_4) \right] 
  \left( \xi_T h^T_{ij} + \xi_L h^L_{ij} \right)\, , 
\eeq
where we have defined $g^{ij}({\bf q}) \equiv g^{ij}({\bf q};\om,{\bf k}) 
= q^i q^j/(\om-{\bf v}_q\cdot {\bf k} + \mi\eps)$ and we have labeled 
the momenta such that ${\bf p}_1={\bf p}$. 
We have also defined the phase space measure $\dif\Gamma_{234}=\dif
\Gamma_1\dif\Gamma_2\dif\Gamma_2$ with $\dif\Gamma_i=\dif^3p_i/(2\pi)^3$ 
and the transition rate $r(1,2;3,4)=w(1,2;3,4)D(1,2;3,4)$. The factor 
$D(1,2;3,4)$ contains the distribution functions 
\beq
D(1,2;3,4) =  f_0({\bf p}_1)f_0({\bf p}_2)
  \left(1-f_0({\bf p}_3)\right)\left(1-f_0({\bf p}_4)\right) \, ,
\eeq
and $w(1,2;3,4)$ is the collision probability
\beq
w(1,2;3,4)\dif\Gamma_{34} = 
   v_{\it rel} \left(\frac{\dif\sigma}{\dif\Omega'}\right) \dif\Omega'\, . 
\eeq
Here, $v_{\it rel}=|{\bf v}_1-{\bf v}_2|$ is the relative velocity, 
$\dif\Omega'$ is the solid angle along $2{\bf q}'={\bf p}_3-{\bf p}_4$
and $(\dif\sigma)/(\dif\Omega')$ is the differential cross section. We 
also define $2\bf{q}=\bf{p}_{1}-\bf{p}_{2}$ where $|\bf{q}|=|\bf{q}'|$.
We will consider a dilute Fermi gas in which the scattering amplitude
is completely characterized by the $s$-wave scattering length $a$. 
The differential cross section is 
\beq
\frac{\dif\sigma}{\dif\Omega'} = \frac{1}{4\pi}
 \frac{a^2}{1+a^2q^2}\, . 
\eeq
In order to solve for $\xi_{T,L}$ we will take moments of the 
linearized Boltzmann equation. The quadratic moments of the 
LHS are given by
\beq 
\label{L_ab}
L^{ab} = \int \dif\Gamma_p\,\frac{\mi \om f_0(1-f_0)}{2mT}
   p^a p^b p^i p^j 
   \sum_{T,L} h^{T,L}_{ij} (\xi_{T,L}+1)\, , 
\eeq
where $L^{ab}=L^{ab}(\om,{\bf k})$. The moments of the RHS are 
given by 
\bea
\label{R_ab}
R^{ab} &=& \frac{\om}{2mT}
 \int\dif\Ga_{tot}\, r(1,2;3,4)\,
   p_1^a p_1^b
  \left[g_{ij}({\bf p}_1)+g^{ij}({\bf p}_2)
          -g^{ij}({\bf p}_3)-g^{ij}({\bf p}_4)\right]
 \sum_{T,L}\xi_{T,L}h_{ij}^{T,L} \nn
&=&  \frac{\om}{2mT}
 \int\dif\Ga_{tot} \, r(1,2;3,4)\,
  \left[p_1^a p_1^b+ p_2^a p_2^b - p_3^a p_3^b - p_4^a p_4^b\right] 
  g^{ij}({\bf p}_1) \sum_{T,L}\xi_{T,L}h_{ij}^{T,L} \nn
&=& \frac{2\om}{m^2T}\int 
  \frac{\dif^3 p \dif^3 q}{(2\pi)^6} \, \dif\Om_{q'}\, 
  \frac{a^2q}{1+a^2q^2} \, D_{2\leftrightarrow 2} \,
    \left( q^a q^b-q'^a q'^b\right) g^{ij}({\bf p}_1)
   \sum_{T,L}\xi_{T,L}h_{ij}^{T,L} \, ,
\eea
with $D_{2\leftrightarrow 2}=D(1,2;3,4)$ and $R^{ab}=R^{ab}(\om,{\bf k})$.

\subsection{Transverse channel}
\label{sec_tr}

 We can now determine $\xi_{T,L}$ by solving $L^{ab}=R^{ab}$. We begin 
by considering $h_{xy} = h_{yx} \neq 0$ with all other components 
of $h_{ij}$ vanishing. In this case $h = 0$ and only $\xi_T$ contributes.
We find 
\bea
L^{xy} &=& \int \dif\Gamma_p\frac{\mi \om f_0(1-f_0)}{2mT}
  \left(\xi_{T}+1\right) p^x p^y p^i p^j h^T_{ij} \nn
&=& \frac{\mi \om \left(\xi_T + 1\right)h_{xy}}{30\pi^2 mT} 
     \int \dif p\, p^6 f_0(1-f_0)
\equiv \frac{\mi\om\left(\xi_T + 1\right)h_{xy}}{15} I_{xy}
\eea
and
\bea
R^{xy} &=& \frac{2\om\xi_T h_{xy}}{m^2T}
  \int \frac{\dif^3 p \dif^3 q}{(2\pi)^6} 
  \,\dif\Om_{q'}\, \frac{a^2q}{1+a^2q^2}\, D_{2\leftrightarrow 2}\, 
  \left( q^x q^y-q'^x q'^y \right) g^{ij}({\bf p}_1)  \nn
&=& \frac{2\om\xi_T h_{xy}}{\pi^3 m T}
  \int \dif p\,\dif q\, \dif\al\,\dif\be\, 
   \frac{a^2 q}{1+a^2 q^2}\, D_{2\leftrightarrow2} \,
   \frac{p^2 q^2 q^x q^y q^x q^y}{2m\om-pk\al-2qk\be + \mi\eps} 
   \equiv \frac{\om\xi_T h_{xy}}{15} C_{xy}\, . 
\eea
Setting $L^{xy} = R^{xy}$ gives 
\beq
\xi_T^{(xy)}(\om,{\bf k}) 
      = -\frac{I_{xy}}{I_{xy} + \mi C_{xy}(\om,{\bf k})}\, .
\eeq
The superscript $(xy)$ indicates that we have obtained $\xi_T$ 
by projecting on the $\Pi^{xy}$ channel. In the limit ${\bf k}
\to 0$ our ansatz \Eqn{del_f_ans} is complete, and there is no 
dependence on the channel. At finite ${\bf k}$, however, additional
tensor structures in \Eqn{del_f_ans} are possible and there is 
some dependence of $\xi_{L,T}$ on the channel. 

 In kinetic theory the stress tensor is given by 
\beq
\label{T_ij_kin}
\de \Pi^{ij} = 2\int \frac{\dif^3p}{(2\pi)^3} \frac{p^i p^j}{m}\de f\, ,
\eeq
where the factor 2 is the spin degeneracy. Using the ansatz 
\Eqn{del_f_ans} we get 
\beq
\de \Pi^{xy} =\frac{\om\xi_T^{(xy)}h_{xy}}{m T}
   \int \frac{\dif^3p}{(2\pi)^3} f_0(1-f_0)
   \frac{p^x p^y p^x p^y}{m\om-p^z k+\mi\eps} 
\equiv \xi_T^{(xy)} h_{xy} J_{xy} \, . 
\eeq
We can now extract the retarded correlation function and the 
spectral function using \Eqn{G_R_lin} and \Eqn{rho_def}. We find
\beq
\label{rho_xy}
\rho^{xyxy}(\om,{\bf k}) = 
    2{\rm Im}\big(\xi_T^{(xy)}(\om,{\bf k})\big)
     {\rm Re}\big(J_{xy}(\om,{\bf k})\big)
  + 2{\rm Re}\big(\xi_T^{(xy)}(\om,{\bf k})\big)
     {\rm Im}\big(J_{xy}(\om,{\bf k})\big)\, . 
\eeq
In general the functions $C_{xy}$ and $J_{xy}$ have to computed 
numerically. The limit ${\bf k}\to 0$ can be studied analytically.
We may consider two cases. In the free case the collision integral
$C_{xy}$ vanishes and $\xi_T=-1$. The spectral function is then 
given by $\rho^{xyxy}(\om,{\bf k})=-2{\rm Im}\,J_{xy}(\om,{\bf k})$.
The function $J_{xy}(\om,{\bf k})$ can be computed analytically 
for all values of $\omega$ and ${\bf k}$. The collisionless
spectral function is 
\beq
\frac{\rho^{xyxy}(\om,{\bf k})}{2\om} = 
 \frac{1}{16\pi}\frac{(2mT)^2}{k}\,
  {\it Li}_2\left( \zeta_F^{-1}\exp\left(
   -\frac{m\omega^2}{2Tk^2}\right)\right) \, , 
\eeq
where ${\it Li}_2(x)$ is a polylogarithm and $\zeta_F=\exp(-\mu/T)$ 
is the fugacity. In the limit ${\bf k}\to 0$ we find
\beq
\label{rho_free}
\lim_{{\bf k}\to 0}\frac{\rho^{xyxy}(\om,{\bf k})}{2\om} 
     = \frac{\varepsilon}{3}\pi\delta(\om)\, , 
\eeq
where $\varepsilon$ is the energy density
\beq
\varepsilon = 2\int \frac{\dif^3 p}{(2\pi)^3} f_0 \frac{p^2}{2m}\ .
\eeq
At finite momentum the delta function is spread out over the regime
$\om\lsim|{\bf k}|(T/m)^{1/2}$. In the collisional case we find that in 
the limit ${\bf k}\to 0$ the collision integral is real and scales as 
$C_{xy}(\om,{\bf k}\to 0)\sim 1/\om$. This means that the spectral function 
is dominated by the ${\rm Im}(\xi^{(xy)}_T(\om,{\bf k})){\rm Re}(J_{xy}
(\om,{\bf k}))$ term. We also find that in this limit ${\rm Re}(J_{xy}) 
\simeq I_{xy}/(15m)$. This implies that 
\beq
\label{eta_om}
 \eta(\om) \equiv \lim_{{\bf k}\to 0}\frac{\rho^{xyxy}(\om,{\bf k})}{2\om} 
  = \frac{I_{xy}^2}{15m\om}
    \frac{{\rm Re}(C_{xy}(\om,{\bf 0}))}
       {I_{xy}^2 + {\rm Re}(C_{xy}(\om,{\bf 0}))^2}\, . 
\eeq
The zero frequency limit is 
\beq
\label{eta_0}
\eta \equiv \lim_{\om\to 0}\eta(\om) 
 = \frac{I_{xy}^2}{15m\om {\rm Re}(C_{xy}(\om,{\bf 0}))} 
 = \frac{15 (mT)^{3/2}}{32 \sqrt{\pi}} 
   \begin{cases}
   1 & a\to \infty \\
   1/(3mTa^2) & a \to 0
   \end{cases}\, , 
\eeq
where in the last step we have evaluated the integrals in the high 
temperature limit. This result agrees with the hydrodynamic limit 
given in \Eqn{G_xy_hydro} and the known formula for the shear viscosity 
of a dilute Fermi gas in kinetic theory \cite{Bruun:2005}. We can 
now write the frequency dependence as
\beq
\eta(\om) = \frac{\eta}{1+\om^2\tau^2}
\eeq
where $\tau = I_{xy}/(\om {\rm Re}(C_{xy})) = 15 m\eta/I_{xy}$ is 
a relaxation time. In the high temperature limit $\tau$ agrees
with the result obtained in \cite{Bruun:2007}. We can also evaluate
the sum rule 
\beq
\label{sum_rule}
\frac{1}{\pi}\int_0^{\infty} \dif \om\, \eta(\om) = \frac{\eta}{2\tau} 
= \frac{I_{xy}}{30m} = \frac{\varepsilon}{3}\, . 
\eeq
We note that the sum rule agrees with the integral of the free 
spectral function given in \Eqn{rho_free}. The sum rule also 
agrees with the general sum rule obtained in \cite{Taylor:2010ju}
if we take account the fact that kinetic theory does not reproduce
the short distance contribution $\eta(\omega)\sim C/\sqrt{m\omega}$,
where $C$ is Tan's contact \cite{Tan:2008}.

\begin{figure}[t]
\includegraphics[width=0.45\textwidth]{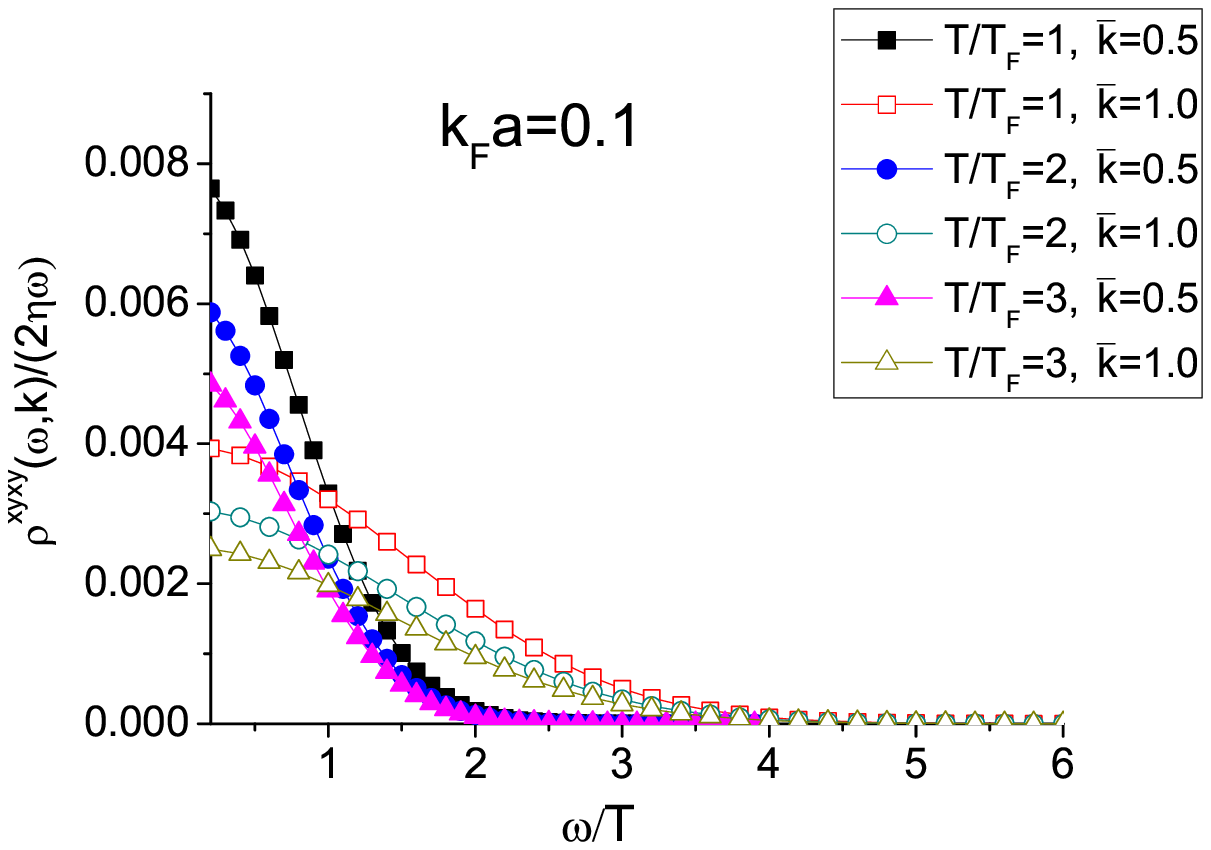}
\includegraphics[width=0.45\textwidth]{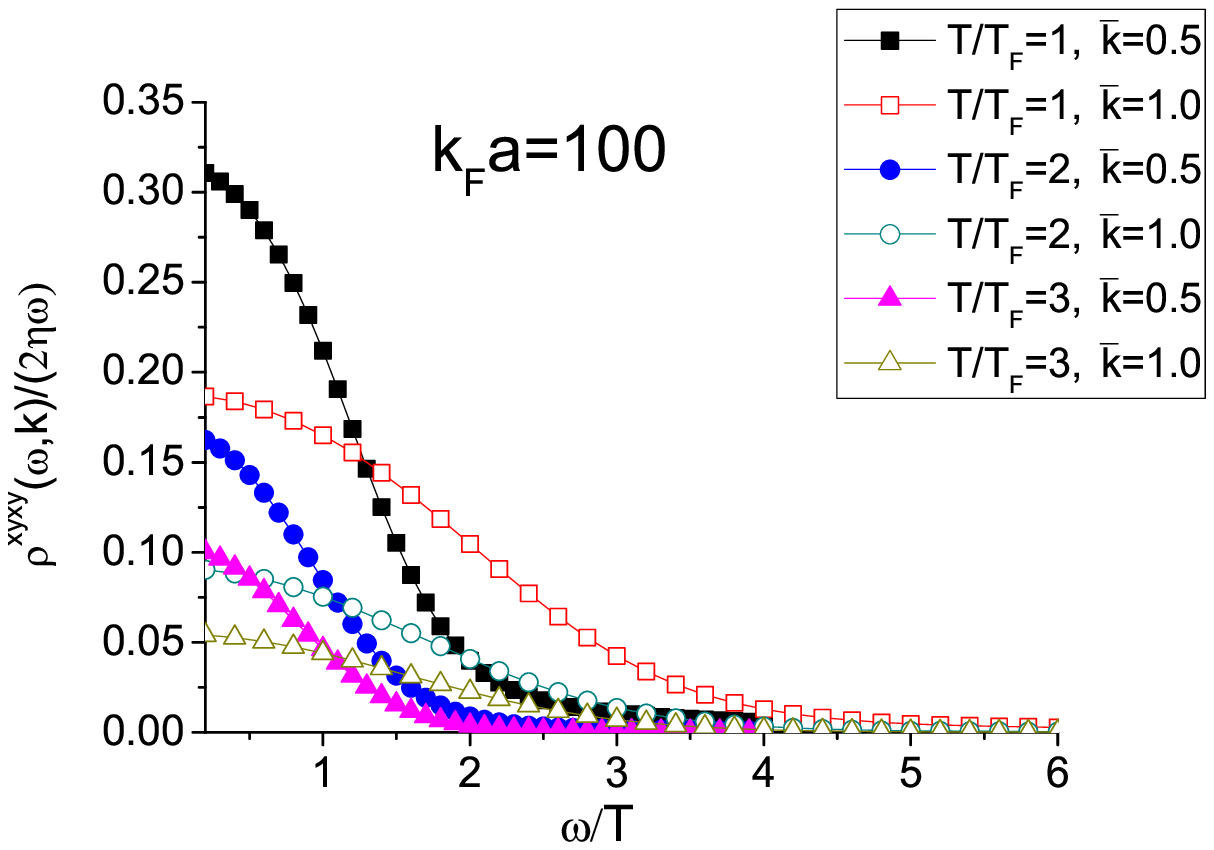}\\
\includegraphics[width=0.45\textwidth]{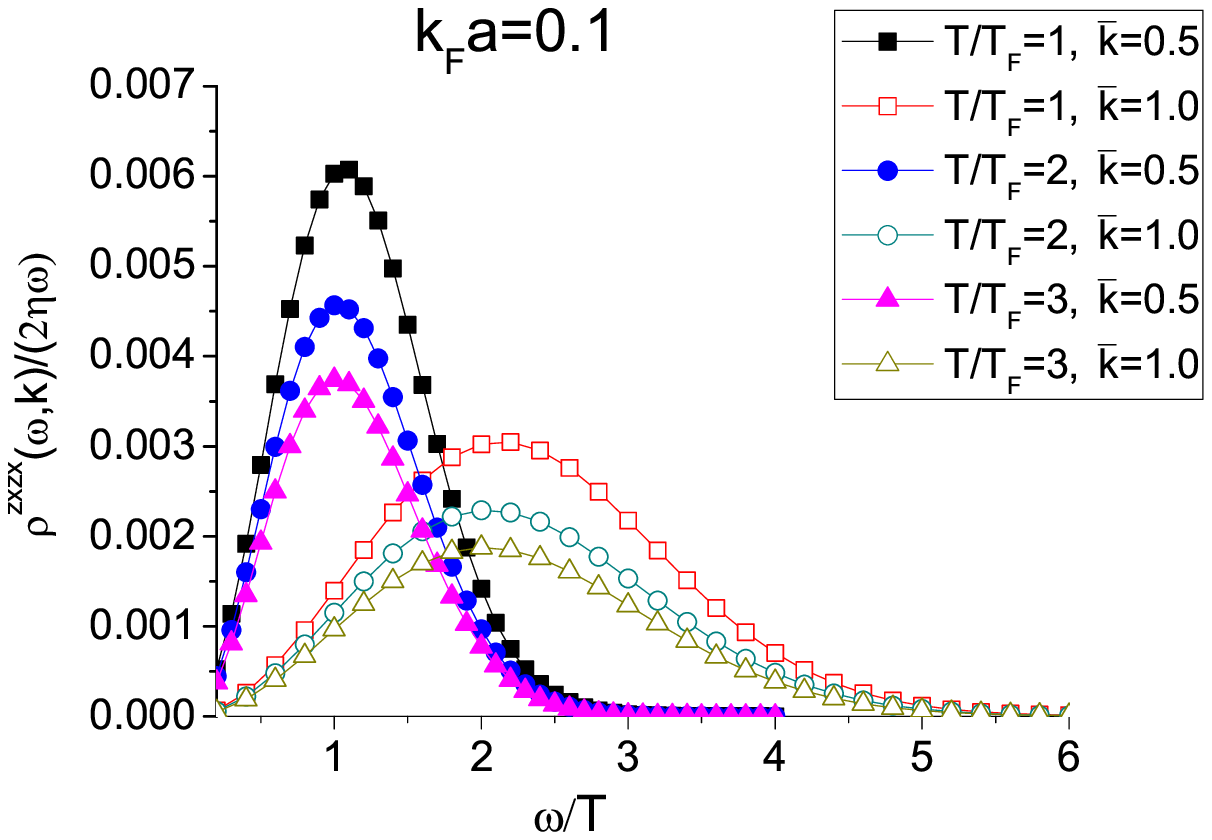}
\includegraphics[width=0.45\textwidth]{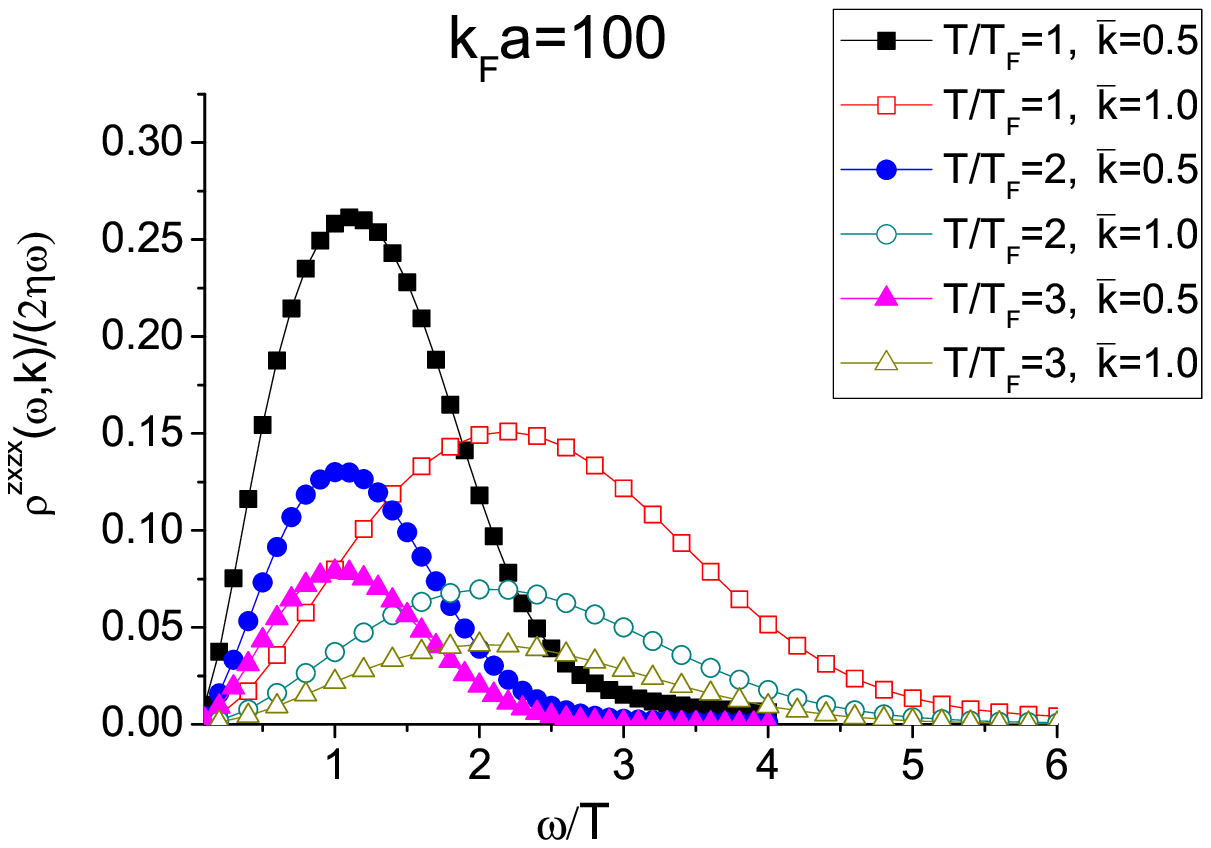}
\caption{Stress tensor spectral functions in the $xy$ and $xz$ channels.
We plot the spectral function $\rho(\omega,k)$ normalized to the 
value in the Kubo limit, $\rho(\omega\to 0,0)=2\eta\omega$, as a 
function of $\omega/T$ for different values of $\bar{k}=k/(2mT)^{1/2}$ 
and $T/T_F$. The panels on the left show the result for $k_Fa=0.1$, 
which corresponds to weak coupling, and the panels on the right 
show the spectral function close to unitarity, for $k_Fa=100$. 
\label{rhot}}
\end{figure}

 The spectral function in the $xz$ channel can be computed in the 
same fashion. The spectral function $\rho^{xzxz}(\om,{\bf k})$ is 
given by \Eqn{rho_xy} with the obvious replacement $(xy)\to (xz)$. 
In the Kubo limit we find
\beq
\label{rho_xz_kubo}
\displaystyle\lim_{\om,{\bf k}\to 0}
   \frac{\rho^{xzxz}(\om,{\bf k})}{2\om} = \eta
\eeq
as expected from hydrodynamics, see \Eqn{G_xz_hydro}. Numerical 
results for the spectral function in the $xy$ and $xz$ channel are 
shown in Fig.~\ref{rhot}. We consider two different scattering length, 
one corresponding to weak coupling, $k_Fa=0.1$, and one close to 
unitarity, $k_Fa=100$. We plot the spectral function as a function 
of $\omega/T$ for several values of $\bar{k}=k/\sqrt{2mT}$ and $T/T_F$.

 We first discuss $\rho^{xyxy}(\om,{\bf k})$. At zero momentum the 
spectral function normalized to $2\eta\omega$ is a Lorentzian with 
a width proportional to $1/\eta$. This implies that the peak in the 
spectral function is narrow in weak coupling, but very broad in strong 
coupling. As the momentum $k$ grows the width of the peak increases
and it becomes less sensitive on the strength of the coupling. In this 
regime the spectral function is close to the free spectral function. 
We also find numerically that the spectral function satisfies the 
sum rule \Eqn{sum_rule} even at non-zero momentum $k$. In the $xz$ 
channel the normalized spectral function $\rho^{xzxz}(\om,{\bf k})
/(2\eta\omega)$ vanishes in the limit $\om\to 0$ for  any non-zero 
value of $k$. For $k=0$ the normalized spectral function approaches 
unity as $\omega\to 0$. This is in agreement with the hydrodynamic 
prediction \Eqn{G_xz_hydro}. We find that the spectral function has 
a peak which, like the peak in the $xy$ channel, corresponds to a 
diffusive mode. The width of the mode is controlled by $k$ except 
in the limit of small momenta where the width is controlled by the 
shear viscosity.

\subsection{Longitudinal case}
\label{sec_long}

  In this section we study the longitudinal spectral function
$\rho^{zzzz}(\omega,{\bf k})$. The calculation is more involved
because both $\xi_{T}$ and $\xi_L$ contribute. We consider a
perturbation $h_{zz} \neq 0$ with all other $h_{ij}=0$. In this
case $h^L_{ij} = \frac{h_{zz}}{3} \de_{ij}$ and $h^T_{zz} = h_{ij}
-\frac{h_{zz}}{3} \de_{ij} = -\frac{h_{zz}}{3}\,{\rm diag}(1,1,-2)$.
The LHS of the Boltzmann equations is $L^{zz}=L^{zz}_T+L^{zz}_L$ 
with 
\beq
\label{left_zz_TL}
L^{zz}_{T,L} = \int \dif\Ga\,\frac{\mi \om f_0 (1-f_0)}{2mT}
  (\xi_{T,L}+1)p^z p^z p^i p^j h_{ij}^{T,L}
   \equiv \frac{\mi\om(\xi_{T,L}+1)h_{zz}}{3}I_{T,L}^{zz} \, .
\eeq
The collision integrals can be written as
\bea
R_T^{zz} &=& \frac{2\om\xi_T}{m^2T}
  \int \frac{\dif^3 p \dif^3 q}{(2\pi)^6} 
       \,\dif \Om_{q'}\, D_{2\leftrightarrow 2}\, 
  \frac{a^2q}{1+a^2q^2} \left(q^z q^z-q'^z q'^z\right) 
   g^{ij}({\bf p_1})h_{ij}^T \nn
 &=& -\frac{\om\xi_T h_{zz}}{12\pi^3 mT}
  \int \dif p\,\dif q\,\dif \al\,\dif\be\,D_{2\leftrightarrow 2}\,
  \frac{a^2q}{1+a^2q^2}\, p^2 q^4\,
  \frac{(\be^2-\third)(p^2(1-3\al^2)-8pq\al\be+4q^2(1-3\be^2))}
  {2m\om-pk\al-2qk\be+\mi\eps} \nn
&\equiv&  \frac{\om \xi_T h_{zz}}{3}C_T^{zz}\, , \\ 
R_L^{zz} &=&\frac{2\om\xi_T}{m^2T}
  \int \frac{\dif^3 p \dif^3 q}{(2\pi)^6} 
       \,\dif \Om_{q'}\, D_{2\leftrightarrow 2}\,
  \frac{a^2q}{1+a^2q^2}\left(q^z q^z-q'^z q'^z\right) 
    g^{ij}({\bf p_1})h_{ij}^L \nn
&=& -\frac{\om\xi_L h_{zz}}{3\pi^3 mT}
  \int\dif p\,\dif q\,\dif \al\,\dif\be\,D_{2\leftrightarrow 2}\,
  \frac{a^2q}{1+a^2q^2}\, p^2 q^4 \, \frac{(\be^2-\third)
  (p^2/4 + pq\al\be + q^2)}{2m\om-pk\al-2qk\be+\mi\eps} \nn
&\equiv&  \frac{\om \xi_L h_{zz}}{3}C_L^{zz} \, . 
\eea
We can now solve for $\xi_{T,L}$. We find
\beq
\xi_{T,L}^{(zz)}(\om,{\bf k}) = -\frac{I_{T,L}^{zz}}
    {I_{T,L}^{zz} + \mi C_{T,L}^{zz}(\om,{\bf k})} \, . 
\eeq
The $zz$ component of the stress tensor is given by 
\bea
\de \Pi^{zz} &=& 2\om\int\frac{\dif^3 p}{(2\pi)^3}\, f_0(1-f_0)
  \frac{p^z p^z}{m}\frac{p^i p^j}{2mT}
  \frac{\xi^{(zz)}_T h^T_{ij} + \xi^{(zz)}_L h^L_{ij}}
       {\om - {\bf v}_p \cdot {\bf k} + \mi\eps} \nn
 &=& \frac{\om h_{zz}}{12\pi^2 mT}
  \int \dif p\,\dif\al\, 
  \frac{f_0(1-f_0)p^6\al^2}{m\om - pk\al+\mi\eps}
  \left[\xi_T^{(zz)}(3\al^2-1) + \xi_L^{(zz)} \right] \nn
  &\equiv & \om h_{zz}
 \left(\xi_T^{(zz)}(\om,{\bf k}) J_T^{zz}(\om,{\bf k}) 
     + \xi_L^{(zz)}(\om,{\bf k}) J_L^{zz}(\om,{\bf k}) \right)\, . 
\eea
We can now extract the retarded correlator and the spectral 
function using \Eqn{G_R_lin} and \Eqn{rho_def}. We get
\beq
\frac{\rho^{zzzz}(\om,{\bf k})}{2\om} = 
   {\rm Im}\left(\xi_T^{(zz)}(\om,{\bf k}) J_T^{zz}(\om,{\bf k}) 
               + \xi_L^{(zz)}(\om,{\bf k}) J_L^{zz}(\om,{\bf k}) \right)\, . 
\eeq
\begin{figure}[t]
\includegraphics[width=0.45\textwidth]{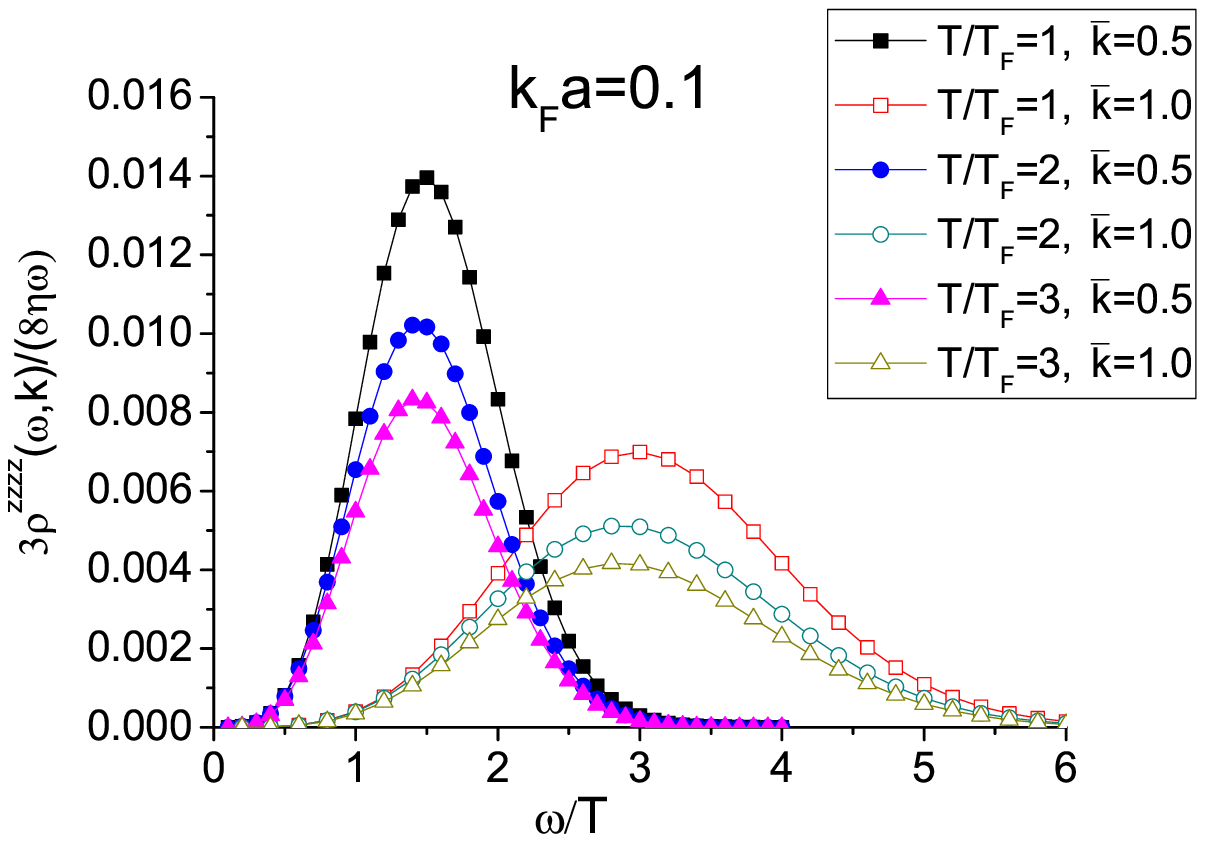}
\includegraphics[width=0.45\textwidth]{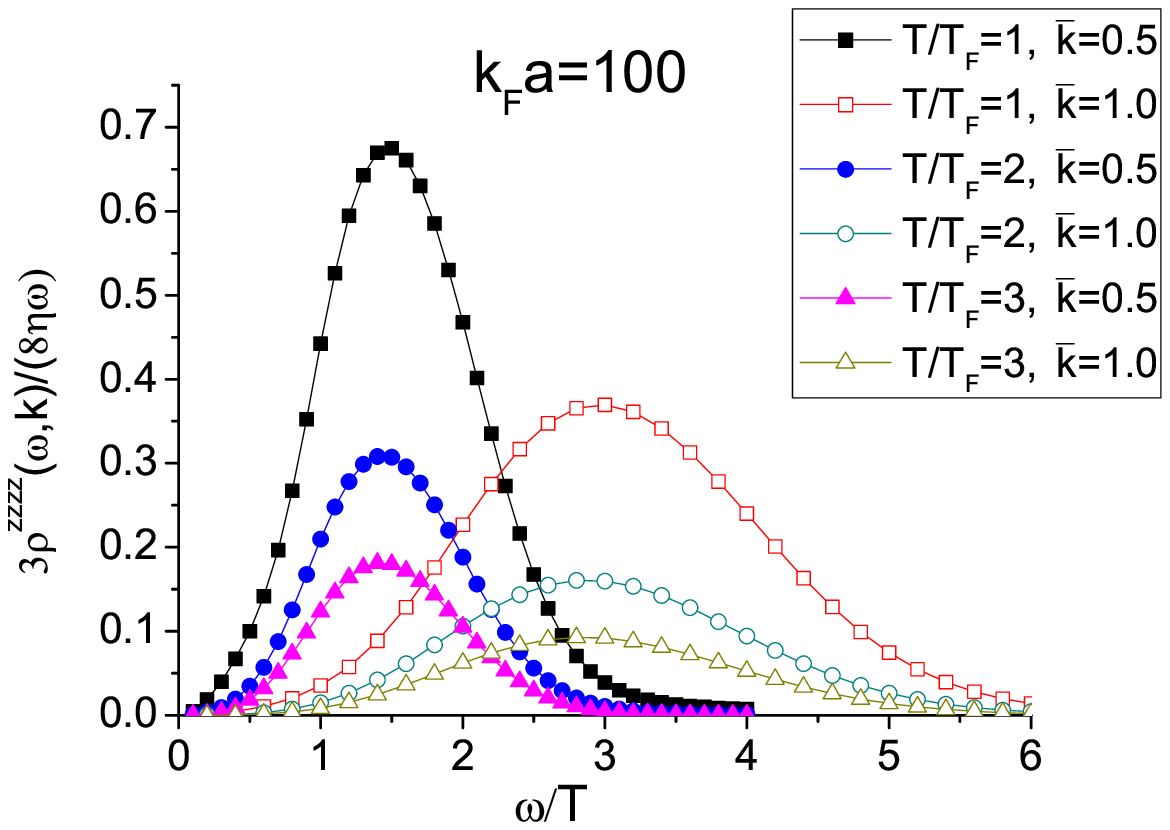}
\caption{Viscosity spectral function in the $zz$ channel.
We plot the spectral function $\rho(\omega,k)$ normalized to the 
value in the Kubo limit, $\rho(\omega\to 0,0)=\frac{8}{3}\eta\omega$, 
as a function of $\omega/T$ for different values of $\bar{k}=k/(2mT)^{1/2}$ 
and $T/T_F$.
\label{rhozz}}
\end{figure}
The result again simplifies in the limit ${\bf k}\to 0$. In the collisional 
case we find that the function $J$ is dominated by its real part. Also, one 
can show that the ${\bf k}\to 0$ limit of ${\rm Re}(C_L^{zz}) = 0$ and 
therefore ${\rm Im}(\xi_L^{(zz)}) = 0$. We find
\beq
\lim_{{\bf k}\to 0}\frac{\rho^{zzzz}(\om,{\bf k})}{2\om} 
  =  \lim_{{\bf k}\to 0}
      {\rm Im} \big(\xi_T^{(zz)}(\om,{\bf k})\big)
      {\rm Re} \big(J_T^{zz}(\om,{\bf k})\big) 
  = \frac{I_T^{zz} {\rm Re}(J_T^{zz}(\om,{\bf 0}))}
         {(I_T^{zz})^2 + {\rm Re}(C_T^{zz}(\om,{\bf 0}))^2}
\eeq
The transverse and longitudinal integrals that we have done are obviously
related to each other. In the limit ${\bf k}\to 0$ we have
\beq
I_T^{zz} = \frac{2}{15} I_{xy} \qquad 
J_T^{zz}(\om,{\bf 0}) = \frac{4}{3}J_{xy}(\om,{\bf 0}) 
             = \frac{4}{45m\om}I_{xy} \qquad 
C_T^{zz} = \frac{2}{15} C_{xy}\, .
\eeq
This shows that as ${\bf k}\to 0$ the ansatz \Eqn{del_f_ans} is complete and 
$\xi_T^{(zz)}(\om)=\xi_T^{(xy)}(\om)$. We can also show that
\beq
\lim_{\om, {\bf k}\to 0}\frac{\rho^{zzzz}(\om,{\bf k})}{2\om} 
 = \frac{4}{3}\eta \, . 
\eeq
Comparing this result with the hydrodynamic prediction (\ref{G_zz_hydro}) 
we find that the bulk viscosity $\zeta$ vanishes. This follows from 
scale invariance at unitarity \cite{Son:2005tj}. The fact that we 
also find $\zeta=0$ away from unitarity is related to the fact that 
we only take into account elastic $2\leftrightarrow 2$ collisions. 
Numerical results for $\rho^{zzzz}(\om,{\bf k})$ normalized to the 
hydrodynamic limit $\frac{8}{3}\omega\eta$ are shown in Fig.~\ref{rhozz}. 
We observe that the spectral function shows a sound peak at $\omega=
c_s k$. At the largest temperature considered, $T/T_F=3$, the speed 
of sound differs from the asymptotic high temperature value $c_s^2=
(5T)/(3m)$ by about 10\%. The width of the sound peak is close
to that of the free spectral function. This is at variance with 
hydrodynamics which predicts that the sound peak is narrow in strong 
coupling, and in the limit of small momentum. At zero momentum 
hydrodynamics predicts that $\rho(\omega)/\omega$ is constant. Higher 
order effects will turn this part of the spectrum into a broad peak
with a width inversely proportional to the relaxation time. This 
implies that at small momentum we expect a narrow peak with 
a width proportional to $(k/c_s)(\eta/\rho)$ superimposed on a broad 
peak with a width proportional $\varepsilon/(T\eta)$. Our ansatz 
correctly reproduces the $k\to 0$ limit as well as the existence
of a sound peak, but it appears to be too simple to reproduce
the two-peak structure at small $k$.

\section{Summary and discussion}
\label{sec_sum}

 We have computed the viscosity spectral function using kinetic 
theory in a non-trivial background metric. 
We find that a simple quadratic ansatz for the non-equilibrium 
distribution function $\delta f$ satisfies the main constraints 
on the spectral function, in particular the existence of a diffusive 
peak and a sound peak (although it misses the interplay between 
these two features in the longitudinal channel), the correct Kubo 
limit, and the viscosity sum rule. Kinetic theory misses the 
non-analytic high frequency tail of the spectral function $\eta
(\omega)\sim C/\sqrt{m\omega}$. We find that the width of the 
diffusive peak at zero momentum is inversely proportional to 
the viscosity. This implies that the relaxation time for the shear 
stress is $\tau \simeq 3\eta/(2\varepsilon)$, in agreement with 
\cite{Bruun:2007}. We also find that at large momentum the spectral 
function is close to the free spectral function, and the width 
of the transport peak is of order $k(T/m)^{1/2}$. 

 There are at least two important questions that we have not 
addressed in this paper. The first is a systematic matching to 
second order hydrodynamics. For this purpose it is important to 
understand the structure of second order hydrodynamics in the 
conformal limit along the lines of the analysis in the relativistic 
case performed in \cite{Baier:2007ix}. The second is a detailed 
study of the imaginary time correlation functions, with an emphasis
on finding the optimal strategy for extracting transport coefficients 
from Quantum Monte Carlo calculations. The main difficulty is that 
the imaginary time correlation function is mainly sensitive to the 
spectral weight of the transport peak, and that the sum rule implies 
that the weight is independent of the shear viscosity. This problem 
can be addressed by utilizing the information contained in the 
momentum dependence of the correlation function, and by considering 
not only the shear channel but also the sound channel 
\cite{Aarts:2002cc,Meyer:2008gt}.

Acknowledgments: This work was supported in parts by the US
Department of Energy grant DE-FG02-03ER41260. We would like
to thank D.~Teaney for useful discussions.


\begin{thebibliography}{99}

\bibitem{Bloch:2007}
I.~Bloch, J.~Dalibard, W~Zwerger
Rev.\ Mod.\ Phys.\ {\bf 80}, 885 (2008)
[arXiv:0704.3011].

\bibitem{Giorgini:2008}
S.~Giorgini, L.~P.~Pitaevskii, S.~Stringari, 
Rev.\ Mod.\ Phys.\ {\bf 80} 1215 (2008)
[arXiv:0706.3360].

\bibitem{Schafer:2009dj}
T.~Sch\"afer and D.~Teaney,
Rept.\ Prog.\ Phys.\  {\bf 72}, 126001 (2009)
[arXiv:0904.3107 [hep-ph]].

\bibitem{oHara:2002}
K.~M.~O'Hara, S.~L.~Hemmer, M.~E.~Gehm, S.~R.~Granade, J.~E.~Thomas,
Science {\bf 298}, 2179 (2002)
[cond-mat/0212463].

\bibitem{Schafer:2007pr}
T.~Sch\"afer,
Phys.\ Rev.\  A {\bf 76}, 063618 (2007)
[arXiv:cond-mat/0701251].

\bibitem{Turlapov:2007}
A.~Turlapov, J.~Kinast, B.~Clancy, L.~Luo, J.~Joseph, J.~E.~Thomas,
J.\ Low Temp.\ Phys.\ {\bf 150}, 567 (2008)
[arXiv:0707.2574].

\bibitem{Schaefer:2009px}
T.~Sch\"afer and C.~Chafin,
arXiv:0912.4236 [cond-mat.quant-gas].

\bibitem{Cao:2010wa}
C.~Cao, E.~Elliott, J.~Joseph, H.~Wu, J.~Petricka, T.~Sch{\"a}fer, 
J.~E.~Thomas, 
preprint [arXiv:1007.2625 [cond-mat.quant-gas]].

\bibitem{Schaefer:2010dv}
T.~Sch{\"a}fer,
preprint [arXiv:1008.3876 [cond-mat.quant-gas]].

\bibitem{Dusling:2007gi}
K.~Dusling and D.~Teaney,
Phys.\ Rev.\  C {\bf 77}, 034905 (2008)
[arXiv:0710.5932 [nucl-th]].

\bibitem{Romatschke:2007mq}
P.~Romatschke and U.~Romatschke,
Phys.\ Rev.\ Lett.\  {\bf 99}, 172301 (2007)
[arXiv:0706.1522 [nucl-th]].

\bibitem{Song:2007ux}
H.~Song and U.~W.~Heinz,
Phys.\ Rev.\  C {\bf 77}, 064901 (2008)
[arXiv:0712.3715 [nucl-th]].

\bibitem{Kovtun:2004de}
P.~Kovtun, D.~T.~Son and A.~O.~Starinets,
Phys.\ Rev.\ Lett.\  {\bf 94}, 111601 (2005)
[arXiv:hep-th/0405231].

\bibitem{Son:2007vk}
D.~T.~Son and A.~O.~Starinets,
Ann.\ Rev.\ Nucl.\ Part.\ Sci.\  {\bf 57}, 95 (2007)
[arXiv:0704.0240 [hep-th]].

\bibitem{Bruun:2007}
G.~M.~Bruun, H.~Smith
Phys. Rev. A {\bf 76}, 045602 (2007) 
[arXiv:0709.1617].

\bibitem{Bruun:2005}
G.~M.~Bruun, H.~Smith,
Phys.\ Rev.\ A {\bf 72}, 043605 (2005) 
[cond-mat/0504734].

\bibitem{Bruun:2006}
G.~M.~Bruun, H.~Smith,
Phys.\ Rev.\ A {\bf 75}, 043612 (2007)
[cond-mat/0612460].

\bibitem{Rupak:2007vp}
G.~Rupak and T.~Sch\"afer,
Phys.\ Rev.\  A {\bf 76}, 053607 (2007)
[arXiv:0707.1520 [cond-mat.other]].

\bibitem{Escobedo:2009bh}
M.~A.~Escobedo, M.~Mannarelli and C.~Manuel,
Phys.\ Rev.\  A {\bf 79}, 063623 (2009)
[arXiv:0904.3023 [cond-mat.quant-gas]].

\bibitem{Braby:2010ec}
M.~Braby, J.~Chao, T.~Sch{\"a}fer,
Phys.\ Rev.\  {\bf A82}, 033619 (2010).
[arXiv:1003.2601 [cond-mat.quant-gas]].

\bibitem{Guo:2010dc}
H.~Guo, D.~Wulin, C.~C.~Chien {\it et al.},
preprint, [arXiv:1008.0423 [cond-mat.quant-gas]].

\bibitem{Enss:2010qh}
T.~Enss, R.~Haussmann, W.~Zwerger,
preprint, [arXiv:1008.0007 [cond-mat.quant-gas]].

\bibitem{Taylor:2010ju}
E.~Taylor and M.~Randeria,
arXiv:1002.0869 [cond-mat.quant-gas].

\bibitem{Nikuni:2004}
T.~Nikuni, A.~Griffin
Phys.\ Rev.\ A {\bf 69}, 023604 (2004)
[arXiv:cond-mat/0309269].

\bibitem{Son:2008ye}
D.~T.~Son,
Phys.\ Rev.\  D {\bf 78}, 046003 (2008)
[arXiv:0804.3972 [hep-th]].

\bibitem{Balasubramanian:2008dm}
K.~Balasubramanian and J.~McGreevy,
Phys.\ Rev.\ Lett.\  {\bf 101}, 061601 (2008)
[arXiv:0804.4053 [hep-th]].

\bibitem{Aarts:2002cc}
G.~Aarts and J.~M.~Martinez Resco,
JHEP {\bf 0204}, 053 (2002)
[arXiv:hep-ph/0203177].

\bibitem{Teaney:2006nc}
D.~Teaney,
Phys.\ Rev.\  D {\bf 74}, 045025 (2006)
[arXiv:hep-ph/0602044].

\bibitem{Moore:2008ws}
G.~D.~Moore and O.~Saremi,
JHEP {\bf 0809}, 015 (2008)
[arXiv:0805.4201 [hep-ph]].

\bibitem{Meyer:2008gt}
H.~B.~Meyer,
JHEP {\bf 0808}, 031 (2008)
[arXiv:0806.3914 [hep-lat]].

\bibitem{Hong:2010}
J.~Hong and D.~Teaney,
arXiv:1003.0699 [nucl-th].

\bibitem{Nishida:2007}
Y.~Nishida and D.~T.~Son, 
Phys.\ Rev.\ D {\bf 76}, 086004 (2007)
[arXiv:0706.3746].

\bibitem{Son:2005rv}
D.~T.~Son and M.~Wingate,
Annals Phys.\  {\bf 321}, 197 (2006)
[arXiv:cond-mat/0509786].

\bibitem{Son:2005tj}
D.~T.~Son,
Phys.\ Rev.\ Lett.\  {\bf 98}, 020604 (2007)
[arXiv:cond-mat/0511721].

\bibitem{Andreasson:2002vz}
H.~Andreasson,
Liv.\ Rev.\ Relativity {\bf 8} (2005)
[gr-qc/0206069].

\bibitem{Bruhat:2009}
Y.~Choquet-Bruhat, 
General Relativity and the Einstein Equations, 
Oxford University Press (2009).

\bibitem{Tan:2008}
S. Tan, 
Annals Phys.\ {\bf 323}, 2952 (2008)
[arXiv:cond-mat/0505200]. 

\bibitem{Baier:2007ix}
R.~Baier, P.~Romatschke, D.~T.~Son {\it et al.},
JHEP {\bf 0804}, 100 (2008).
[arXiv:0712.2451 [hep-th]].


\end{thebibliography}
\end{document}